\documentstyle[preprint,pre,eqsecnum,aps]{revtex}
\begin{document}
\draft
\title{Periodicity of cylindrical linear cellular automata}
\author{Shin-ichi Tadaki\cite{EADDRESS}}
\address{Department of Information Science,
Saga University, Saga 840, Japan}
\date{\today}
\maketitle
\begin{abstract}
Periodicity and relaxation are investigated for the
trajectories of the states in cylindrical linear cellular
automata.  The time evolutions are described with matrices.
The eigenvalue analysis is applied to obtain the maximum
values of period and relaxation.  The translational
invariance suppresses the maximum period and relaxation.
\end{abstract}
\narrowtext
\section{Introduction}
Cellular automata (CA) are one of the simplest mathematical
models for nonlinear dynamics to produce complex patterns of
behavior.  They had been originally introduced by von
Neumann\cite{Neumann} to investigate some artificial life.
Wolfram had reintroduced CA as a model to
investigate complexity and randomness\cite{Wolfram:1}. He
investigated many fundamental features of
them\cite{Wolfram:2,Wolfram:3,Martin}.
Since then many authors have made efforts to clarify the
properties of CA and applied to natural
systems including biological systems, complex fluids, chemical
reactions, astronomical systems and so on\cite{CA}.

Computational simplicity is the advantages of using CA to
simulate complex behaviors. Discrete states are defined on
each lattice point and evolve with discrete time steps in
CA.  Moreover nonlinear properties are expected to be
emphasized by discretization.  A simple modeling, however,
does not mean simple properties.  Even though the great
efforts, the fundamental properties of CA are still unclear.
Studies on these simple systems seems to be a key for
understanding discretized simulated systems.

One-dimensional CA are described by the discrete time
evolution of site \(a_i\):
\begin{equation}
a_i(t+1) = F[a_{i-r}(t),
a_{i-r+1}(t),\ldots,a_i(t),\ldots,a_{i+r}(t)],
\end{equation}
where $a_i$ takes $k$ discrete values over $Z_k$.  The
simplest models, {\em elementary cellular automata}, consist
of sites with two internal states over $Z_2$ ($k=2$)
interacting with the nearest neighbor sites ($r=1$).
Wolfram introduced a naming scheme for these models and
classified the behavior of CA into four
classes\cite{Wolfram:1,Wolfram:2}.

Most authors have worked on CA within the scope of the
infinite number of sites.  A few works have concerned the
effects of finiteness.  The orbits of {\it cylindrical}
linear CA, which are CA with periodic boundaries, are
analyzed on the basis of {\it characteristic polynomials},
which directly describe the states of linear
CA\cite{Martin,Wolfram:4,Jen,Voorhees,Guan}.  Stevens,
Rosensweig and Cerkanowicz had investigated rule 90 CA,
which will be described in \ref{Model}, with Dirichlet
boundary conditions.  They analyzed the eigenvalue
polynomials of the matrices which give the time evolution of
the system\cite{Stevens}.  In our previous
papers\cite{Tadaki:1,Tadaki:2} (referred as papers I and
II), we had also investigated the periodic orbits of finite
linear CA (rule 90 and 150) with Dirichlet boundary
conditions by analyzing the eigenvalue equations. In the
present paper the method is applied to cylindrical linear
CA. The proof of the classification of the orbits for
Dirichlet boundary cases is given in Appendix
\ref{Appendix-Classify}.

\section{The models and orbits of states}\label{Model}
There are two examples of linear cellular automata in the
elementary ones (\(k=2, r=1\)).  They are called as rule 90
and rule 150 following the Wolfram's naming scheme.  For
rule 90 CA, the time evolution of the $i$-th site
$a_i\in\{0,1\}$($i=1,\ldots,N$) is described as a sum modulo
2 of the nearest-neighbor sites:
\begin{equation}
a_i(t+1)=a_{i-1}(t)+a_{i+1}(t)\bmod2.
\end{equation}
The time evolution of rule 150 CA is given as a sum modulo 2
of the nearest-neighbor sites and itself:
\begin{equation}
a_i(t+1)=a_{i-1}(t)+a_i(t)+a_{i+1}(t)\bmod2.
\end{equation}
We use the periodic boundary conditions: $a_0=a_N,
a_{N+1}=a_1$.  These models belong to the third class which
show the chaotic behavior in the Wolfram's classification

These models are linear because the rules are additive and
the time evolutions are also expressed by the matrices
\begin{equation}
A(t+1)=U A(t),
\end{equation}
where $A(t) =\ ^t(a_1(t),a_2(t),\ldots,a_N(t))$ describes the
state, $N$ bits binary number, at $t$.
The components of the transfer matrices $U$ are given by
\begin{eqnarray}
U_{ij} &=&\cases{1&$j=i\pm1$\cr
                 1&$i=1,\ j=N$\cr
                 1&$i=N,\ j=1$\cr
                 0&otherwise\cr}\quad\hbox{(rule-90)},\\
       &=&\cases{1&$j=i\pm1$\cr
                 1&$j=i$\cr
                 1&$i=1,\ j=N$\cr
                 1&$i=N,\ j=1$\cr
                 0&otherwise\cr}\quad\hbox{(rule-150)}.
\end{eqnarray}

The trajectories of the states can be classified into three
cases with the properties of the transfer matrices as
mentioned in papers I and II (See Fig.~\ref{bigschematic}).
\begin{eqnarray}
U^{\Pi_N}&=&I,\label{periodic}\\
U^{\Pi_N+\pi_N}&=&U^{\pi_N},\label{relaxation}\\
U^{f_N}&=&0,\label{null}
\end{eqnarray}
where $I$ denotes a unit matrix.  The first case,
Eq.~(\ref{periodic}), corresponds to perfectly periodic
motions with the maximum period $\Pi_N$.  Namely all states
belong to perfectly periodic orbits whose period does not
exceed $\Pi_N$.  The second case, Eq.~(\ref{relaxation}),
shows periodic motions with relaxation.  A relaxation path
whose time step is not greater than $\pi_N$ leads to a
periodic orbit with the maximum period $\Pi_N$.  The third,
Eq.~(\ref{null}), is a special case of the second one. All
states are drawn to a null state on the contrary to that the
current CA belong to the third class of Wolfram's
classification of motions (chaotic motions).

\section{Eigenvalue analysis}
The eigenvalue polynomials for $N$-site linear cellular
automata are defined by
\begin{equation}
D^{\rm(R)}_N(\lambda) = \vert U+\lambda I\vert,
\end{equation}
where the index R specifies the rule number and the boundary
condition, for example, 90D for rule 90 CA with Dirichlet
boundaries and 150P for rule 150 with periodic boundaries.
Since each site $a_i$ takes binary values, the eigenvalue
polynomials are over $Z_2$, namely each coefficient is 0 or
1. The eigenvalues $\lambda$ are not usual numbers but over
the Galois field GF($2^N$), a finite field with $2^N$
elements.

The eigenvalue polynomials enable us to find the maximum
period $\Pi_N$ and the maximum relaxation $\pi_N$ as
discussed in Ref. \cite{Stevens}, papers I and II.  For
nilpotent cases, $D^{\rm(R)}_N(\lambda)=\lambda^N$, all
states are drawn into a null state within $N$ steps or less.
If the eigenvalue polynomial has a constant term,
$D^{\rm(R)}_N(0)=1$, the eigenvalue equation
$D^{\rm(R)}_N(\lambda)=0$ can be reduced to a simple form as
$\lambda^{P_N}+1=0$ by multiplying some powers of $\lambda$
and repeatedly substituting the eigenvalue
equation\cite{num}.  Then the minimum value of $P_N$
corresponds to the maximum period $\Pi_N$.  In other words
the maximum period $\Pi_N$ is the minimum integer $m$
satisfying $D^{\rm(R)}_N(\lambda)\vert(\lambda^m-1)$, where
the notation $f\vert g$ denotes $f$ divides $g$.  The
remaining cases are mixture of above two cases as
$D^{\rm(R)}_N(\lambda)=\lambda^{p_N}\times d(\lambda)$,
where the polynomial $d(\lambda)$ has a constant term ($
D^{\rm(R)}_N(0)=0$ and $d(0)=1$). One can evaluate the
maximum period with the above procedure from $d(\lambda)$.
The maximum period is the minimum integer $m$ satisfying
$d(\lambda)\vert(\lambda^m-1)$.  The relaxation length to
periodic orbits is $p_N$.  Actual periods and relaxation
paths are given by divisors of the maximum values
respectively.  They depend on the symmetry of the initial
states as shown in papers I and II.

The eigenvalue polynomials for rule 90 cylindrical CA obey
the relation
\begin{equation}
D_N^{\rm(90P)}(\lambda)=\lambda D_{N-1}^{\rm(90D)}(\lambda).
\label{REL90}
\end{equation}
Orbits of rule 90 cylindrical CA can be
classified as
\begin{eqnarray}
&U^{\Pi_N+1}=U&(N\hbox{ is odd}),\\
&U^{\Pi_N+\pi_N}=U^{\pi_N}\quad&(N\hbox{ is even except }N=2^n),\\
&U^{N}=0&(N=2^n),
\end{eqnarray}
by Eq. (\ref{REL90}) and the classification of orbits for
rule 90 finite CA with Dirichlet boundaries (see Appendix
\ref{Appendix-Classify}).  There are no perfect periodic
orbits by comparing with the Dirichlet boundary cases.

The maximum values of the period and relaxation are derived
from the polynomials
\begin{equation}
D_N^{\rm(90P)}(\lambda) =
\sum_{j=0}^{\lfloor (N-1)/2\rfloor}
\left(C^{N-1}_j\bmod 2\right) \lambda^{N-2j},
\end{equation}
which are the solutions of Eq. (\ref{REL90}), where $\lfloor
x\rfloor$ denotes the largest integer not exceeding $x$ and
\begin{equation}
C^N_j = (-1)^j{N-j\choose j}.
\end{equation}
Some examples of the eigenvalue polynomials are shown in
Table~\ref{poly90}.  Equation (\ref{REL90}) leads that the
maximum period for rule 90 cylindrical CA with $N$ sites are
equal to that for $N-1$-site case with Dirichlet boundaries,
$\Pi_N ({\rm rule\ 90P}) = \Pi_{N-1} ({\rm rule\ 90D})$.

The eigenvalue polynomials for rule 150 cylindrical CA can
be also written with those with Dirichlet boundaries
\begin{equation}
D_N^{\rm(150P)}(\lambda)=(1+\lambda)D_{N-1}^{\rm(150D)}(\lambda).
\label{REL150}
\end{equation}
Orbits can be classified as
\begin{eqnarray}
&U^{\Pi_N}=I&(N\not=3n),\\
&U^{N}=I&(N=2^n),\\
&U^{\Pi_N+\pi_N}=U^{\pi_N}\quad&(N=3n).
\end{eqnarray}
by virtue of the classification of orbits for rule 150 CA
with Dirichlet boundaries.  There are no characteristic
difference from the Dirichlet boundary cases.  No simple
relations on the maximum periods between the periodic and
Dirichlet boundary cases are found.

The maximum periods and relaxations are derived from the
polynomials
\begin{equation}
D_N^{\rm(150P)}(\lambda) =
\left(\sum_{j=0}^{\lfloor (N-1)/2 \rfloor}C^{N-1}_j\bmod2\right)
+\sum_{k=1}^N
\left(\sum_{j=0}^{\lfloor (N-k)/2\rfloor}
C^{N-1}_j {N-2j\choose k}\bmod2\right)
\lambda^k.
\end{equation}
Some examples of the eigenvalue polynomials are shown in
Table~\ref{poly150}.

The actual period and relaxation of the system may be
suppressed by the translational symmetry which the periodic
boundary conditions ensure.  Direct matrix multiplications
are also the tool to find the maximum period and relaxation.
The results are summarized in Table \ref{actual-period} with
those obtained from the eigenvalue analysis.  The suppressed
maximum period, which means the maximum period obtained by
the direct matrix multiplications, are just a half of those
by the eigenvalue analysis.  On the relaxation, the same
situations happen except for the $\pi_N=1$ cases.  The
suppressed maximum periods are plotted in
Figs.~\ref{period90} and \ref{period150}.

As the results from the suppression of the maximum values of
the period and relaxation, we can find a simple relation of
the period of rule 90 CA with periodic and Dirichlet
boundaries.  The period of $N$-site rule 90 cylindrical CA
is half of $N-1$-site rule 90 CA with Dirichlet boundaries.
No such simple relations are found for rule 150 cases.

\section{Summary and Discussion}
The method of the eigenvalue analysis was applied to linear
cellular automata with periodic boundaries.  The
trajectories of the states of CA were classified.  The
maximum values of the period and relaxation were evaluated
by the eigenvalue equations.  For rule 90 cases a simple
relation was found between the eigenvalue polynomials with
periodic boundaries and those with Dirichlet boundaries.

The cylindrical CA enjoy the translational
invariance with periodic boundaries.  This translational
symmetry suppresses the maximum values of the period and
relaxation.

In the previous papers\cite{Tadaki:1,Tadaki:2} I had found
the symmetry of the initial states suppress the period.  So
it is natural that the translational symmetry suppresses the
period and relaxation.  The translational symmetry changes
the form of the transfer matrices $U$ and consequently the
form of the eigenvalue polynomials.  The eigenvalue
polynomials, however, do not tell us the suppression of the
period and relaxation.

The eigenvalue analysis may not be able to handle the
translational invariance well.  The matrix $U$ for rule 90
CA can be decomposed as $U=L+R$, where $L$ and $R$ denote
the translation operation to the left and right direction
respectively.  For rule 150 CA it can be describe as
$U=L+R+I$. Namely the dynamics is described only with the
composition of the translation operation.  This may be the
reason of the different results from the eigenvalue analysis
and the direct matrix multiplications.

Let me investigate the cas of $N=4$ rule 90 CA with periodic
boundary conditions for instance.
The eigenvalue equation is $\lambda^4=0$. This can be expressed with the
transfer matrix $U$ as
\begin{equation}
U^4=(L+R)^4=L^4+R^4=I+I=0,
\end{equation}
where all operations are carried over GF(2).
On the other hand, another identity
\begin{equation}
U^2=(L+R)^2=L^2+R^2=0
\end{equation}
holds by virtue of the identity
\begin{equation}
L^m=R^{N-m}.
\end{equation}
This identity may not be taken account into the eigenvalue
analysis.

\acknowledgements
The author would like to thank S. Matsufuji, K. Niizeki and
S. Takagi for valuable discussions and comments.

\appendix
\section{Classification of orbits for Dirichlet boundary cases}
\label{Appendix-Classify}
I had given the classification of orbits of rule 90 and rule
150 finite linear cellular automata with Dirichlet
boundaries in paper II. Here I give brief proofs of the
classification.

For rule 90 cases, the eigenvalue polynomials obey the
following recursion relation:
\begin{equation}
D^{(\rm 90D)}_N(\lambda) = \lambda D^{(\rm 90D)}_{N-1}(\lambda)
- D^{(\rm 90D)}_{N-2}(\lambda).
\label{REQ90}
\end{equation}
The equation (\ref{REQ90}) is simplified for $\lambda=0$
case to $D^{(\rm 90D)}_N(0) = D^{(\rm 90D)}_{N-2}(0)$.  With
the special values $D^{(\rm 90D)}_3(0)=0$ and
$D^{(\rm 90D)}_4(0)=1$, one can find simple relations:
\begin{eqnarray}
D^{(\rm 90D)}_{2n}(0)&=&D^{(\rm 90D)}_{2n-2}(0)=D^{(\rm 90D)}_4(0)=\cdots=1,
              \label{PP90}\\
D^{(\rm 90D)}_{2n+1}(0)&=&D^{(\rm 90D)}_{2n-1}(0)=D^{(\rm 90D)}_3(0)=\cdots=0.
              \label{IP90}
\end{eqnarray}
The eigenvalue polynomials for even $N$ have a
constant term (Eq. (\ref{PP90})). This means that rule 90
CA with Dirichlet boundaries with even number
of sites show perfectly periodic motions.  On the contrary,
Eq. (\ref{IP90}) shows that rule 90 CA with
Dirichlet boundaries with odd number of sites have periodic
orbits with relaxation paths or paths drawn into a null
state.

The another recursion relation
\begin{equation}
D^{(\rm 90D)}_N(\lambda)=\lambda^2D^{(\rm 90D)}_{N-2}(\lambda)
+ D^{(\rm 90D)}_{N-4}(\lambda)\label{REQ90N}
\end{equation}
is derived from Eq. (\ref{REQ90}).
By solving Eq. (\ref{REQ90N}) with
$D^{(\rm 90D)}_{3}(\lambda)=\lambda^3$ and
$D^{(\rm 90D)}_{5}(\lambda)=\lambda^5+\lambda$,
a simple relation
\begin{equation}
D^{(\rm 90D)}_{2n+1}(\lambda)=\lambda D^{(\rm 90D)}_{n}(\lambda^2)
\label{REQ90NN}
\end{equation}
is derived\cite{errata}.
This gives explicit expressions of the eigenvalue polynomials for
$N=2^n-1$ cases:
\begin{equation}
D^{(\rm 90D)}_{2^n-1}(\lambda)=\lambda D^{(\rm 90D)}_{2^{n-1}-1}(\lambda^2)=
\cdots = \lambda^{2^n-1}.\label{EXP90}
\end{equation}
On the other hand, Eq. (\ref{REQ90}) gives the explicit expressions of the
eigenvalue polynomials with binomial coefficients as
\begin{equation}
D^{(\rm 90D)}_{2^n-1}(\lambda)=
\sum_{J=0}^{2^{n-1}}\left({2^n-1-j\choose j}\bmod 2\right)
\lambda^{2^n-1-2j}.\label{EXP90B}
\end{equation}
Comparing Eqs. (\ref{EXP90}) and (\ref{EXP90B}), identities
on binomial coefficients
\begin{equation}
{2^n-1-j\choose j}\bmod 2=0\ (j\not=0)\label{BIID}
\end{equation}
are obtained.  The eigenvalue polynomials for odd $N(\not=2^n-1)$ can
be expressed with those for even $N$ by repeated usages of
Eq. (\ref{REQ90NN}).
So they can not be nilpotent.
Therefore the following classification of orbits
\begin{eqnarray}
&U^{\Pi_N}=I&(N\hbox{ is even}),\\
&U^{\Pi_N+\pi_N}=U^{\pi_N}&(N\hbox{ is odd}(\not=2^n-1)),\\
&U^{N}=0&(N=2^n-1),
\end{eqnarray}
is proven.

For rule 150 cases, the recursion relation changes to
\begin{equation}
D^{(\rm 150D)}_N(\lambda) = (1+\lambda) D^{(\rm 150D)}_{N-1}(\lambda)
- D^{(\rm 150D)}_{N-2}(\lambda).
\label{REQ150}
\end{equation}
This leads to a simple relation $D^{(\rm 150D)}_N(0)=D^{(\rm 150D)}_{N-3}(0)$
with $D^{(\rm 150D)}_3(0)=D^{(\rm 150D)}_4(0)=1$ and
$D^{(\rm 150D)}_5(0)=0$. Therefore one canobtain simple relations
\begin{eqnarray}
D^{(\rm 150D)}_{3n}(0)&=\cdots=&D^{(\rm 150D)}_{3}(0)=1,\\
D^{(\rm 150D)}_{3n+1}(0)&=\cdots=&D^{(\rm 150D)}_{4}(0)=1,\\
D^{(\rm 150D)}_{3n+2}(0)&=\cdots=&D^{(\rm 150D)}_{5}(0)=0.
\end{eqnarray}
Orbits of rule 150 finite linear CA
are perfectly periodic except $N=3n+2$ cases.

One can prove that the eigenvalue polynomials of rule 150
CA with $N=3n+2$ are not nilpotent for $n>0$.
Corresponding to Eqs. (\ref{REQ90N}) and (\ref{REQ90NN}),
recursion relations
\begin{eqnarray}
D^{(\rm 150D)}_N(\lambda)&=&(1+\lambda^2)D^{(\rm 150D)}_{N-2}(\lambda)
                          +D^{(\rm 150D)}_{N-4}(\lambda),\\
D^{(\rm 150D)}_{2n+1}(\lambda)&=&(1+\lambda)D^{(\rm 150D)}_{n}(\lambda^2),
\end{eqnarray}
hold. Applying them to odd $3n+2$, namely odd $n$,  one can obtain
the new recursion relation
\begin{equation}
D^{(\rm 150D)}_{3n+2}(\lambda)=
(1+\lambda)D^{(\rm 150D)}_{3((n-1)/2)+2}(\lambda^2).\label{REQ150NN}
\end{equation}
By the special case $D^{(\rm 150D)}_{5}(\lambda)=\lambda^5+\lambda^4$,
$D^{(\rm 150D)}_{3n+2}(\lambda)$ can not be nilpotent for odd $n$.
For the even $3n+2$ cases, the new recursion relation
\begin{equation}
D^{(\rm 150D)}_{N}(\lambda)=\lambda^4D^{(\rm 150D)}_{N-4}(\lambda)
+(1+\lambda^2)D^{(\rm 150D)}_{N-6}(\lambda)
\end{equation}
is used. The lowest order term of $D^{(\rm
150D)}_{8}(\lambda)=\lambda^8+\lambda^6+\lambda^2$ is
$\lambda^2$. And if $D^{(\rm 150D)}_{3n+2}(\lambda)$ has a $\lambda^2$
as the lowest order,
\begin{equation}
D^{(\rm 150D)}_{3n+8}(\lambda)=\lambda^4D^{(\rm 150D)}_{3n+4}(\lambda)
+(1+\lambda^2)D^{(\rm 150D)}_{3n+2}(\lambda)
\end{equation}
tells that also the lowest order of $D^{(\rm 150D)}_{3n+8}(\lambda)$
is $\lambda^2$. Namely $D^{(\rm 150D)}_{3n+2}(\lambda)$
must have $\lambda^{3n+2}$ and $\lambda^2$ terms at least and
can not be
nilpotent for even $3n+2(n>2)$.

Let me investigate special cases with $N=2^n-1$.
The recursion relation Eq. (\ref{REQ150}) gives
\begin{equation}
D^{(\rm 150D)}_N(\lambda)=\sum_{k=0}^N
\left(\sum_{j=0}^{\lfloor (N-k)/2\rfloor}
{N-j\choose j}{N-2j\choose k}\bmod2\right) \lambda^k.
\end{equation}
For special cases $N=2^n-1$, this reduces to
\begin{equation}
D^{(\rm 150D)}_{2^n-1}(\lambda)=\sum_{k=0}^N \lambda^k,
\end{equation}
with the identity Eq. (\ref{BIID}) and ${2^n-1\choose
k}\bmod2=1(2^n-1\ge k\ge 0)$.  This gives explicit
expressions of the periods as $\Pi_{2^n-1}=2^n$.
Therefore the classification of orbits for rule 150
CA with Dirichlet boundaries
\begin{eqnarray}
&U^{\Pi_N}=I&(N\not=3n+2),\\
&U^{N}=I&(N=2^n-1),\\
&U^{\Pi_N+\pi_N}=U^{\pi_N}&(N=3n+2),
\end{eqnarray}
is proven for $N>2$.

\begin{figure}
\caption{Schematic features of the trajectories of cellular automata:
(a) simple periodic orbit, (b) periodic orbit with relaxation,
(c) limit point.}
\label{bigschematic}
\end{figure}

\begin{figure}
\caption{Actual period for rule 90 cylindrical cellular automata}
\label{period90}
\end{figure}

\begin{figure}
\caption{Actual period for rule 150 cylindrical cellular automata}
\label{period150}
\end{figure}


\widetext
\begin{table}
\caption{Eigenvalue polynomials of rule 90 cylindrical cellular automata}
\label{poly90}
\begin{tabular}{rl}
4 &$ \lambda^{4}$\\
5 &$ \lambda^{5} + \lambda^{3} + \lambda^{1}$\\
6 &$ \lambda^{6} + \lambda^{2}$\\
7 &$ \lambda^{7} + \lambda^{5} + \lambda^{1}$\\
8 &$ \lambda^{8}$\\
9 &$ \lambda^{9} + \lambda^{7} + \lambda^{5} + \lambda^{1}$\\
10 &$ \lambda^{10} + \lambda^{6} + \lambda^{2}$\\
11 &$ \lambda^{11} + \lambda^{9} + \lambda^{5} + \lambda^{3} + \lambda^{1}$\\
12 &$ \lambda^{12} + \lambda^{4}$\\
13 &$ \lambda^{13} + \lambda^{11} + \lambda^{9} + \lambda^{3} + \lambda^{1}$\\
14 &$ \lambda^{14} + \lambda^{10} + \lambda^{2}$\\
15 &$ \lambda^{15} + \lambda^{13} + \lambda^{9} + \lambda^{1}$\\
16 &$ \lambda^{16}$\\
17 &$ \lambda^{17} + \lambda^{15} + \lambda^{13} + \lambda^{9} + \lambda^{1}$\\
18 &$ \lambda^{18} + \lambda^{14} + \lambda^{10} + \lambda^{2}$\\
19 &$ \lambda^{19} + \lambda^{17} + \lambda^{13} + \lambda^{11}
     + \lambda^{9} + \lambda^{3} + \lambda^{1}$\\
20 &$ \lambda^{20} + \lambda^{12} + \lambda^{4}$\\
21 &$ \lambda^{21} + \lambda^{19} + \lambda^{17} + \lambda^{11}
     + \lambda^{9} + \lambda^{5} + \lambda^{3} + \lambda^{1}$\\
22 &$ \lambda^{22} + \lambda^{18} + \lambda^{10} + \lambda^{6} + \lambda^{2}$\\
23 &$ \lambda^{23} + \lambda^{21} + \lambda^{17} + \lambda^{9}
     + \lambda^{7} + \lambda^{5} + \lambda^{1}$\\
24 &$ \lambda^{24} + \lambda^{8}$\\
25 &$ \lambda^{25} + \lambda^{23} + \lambda^{21} + \lambda^{17}
     + \lambda^{7} + \lambda^{5} + \lambda^{1}$\\
26 &$ \lambda^{26} + \lambda^{22} + \lambda^{18} + \lambda^{6} + \lambda^{2}$\\
27 &$\lambda^{27} + \lambda^{25} + \lambda^{21} + \lambda^{19}
     + \lambda^{17} + \lambda^{5} + \lambda^{3} + \lambda^{1}$\\
28 &$ \lambda^{28} + \lambda^{20} + \lambda^{4}$\\
29 &$\lambda^{29} + \lambda^{27} + \lambda^{25} + \lambda^{19}
     + \lambda^{17} + \lambda^{3} + \lambda^{1}$\\
30 &$ \lambda^{30} + \lambda^{26} + \lambda^{18} + \lambda^{2}$\\
31 &$ \lambda^{31} + \lambda^{29} + \lambda^{25} + \lambda^{17}
     + \lambda^{1}$\\
32 &$ \lambda^{32}$\\
\end{tabular}
\end{table}

\begin{table}
\caption{Eigenvalue polynomials of rule 150 cylindrical cellular automata}
\label{poly150}
\begin{tabular}{rl}
3 &$\lambda^{3} + \lambda^{2}$\\
4 &$\lambda^{4} + 1$\\
5 &$\lambda^{5} + \lambda^{4} + \lambda^{3} + \lambda^{2} + \lambda^{1} + 1$\\
6 &$\lambda^{6} + \lambda^{4}$\\
7 &$\lambda^{7} + \lambda^{6} + \lambda^{3} + \lambda^{2} + \lambda^{1} + 1$\\
8 &$\lambda^{8} + 1$\\
9 &$\lambda^{9} + \lambda^{8} + \lambda^{7} + \lambda^{6} + \lambda^{3}
   + \lambda^{2}$\\
10 &$\lambda^{10} + \lambda^{8} + \lambda^{6} + \lambda^{4} + \lambda^{2} +
1$\\
11 &$\lambda^{11} + \lambda^{10} + \lambda^{5} + \lambda^{4} + \lambda^{1} +
1$\\
12 &$\lambda^{12} + \lambda^{8}$\\
13 &$\lambda^{13} + \lambda^{12} + \lambda^{11} + \lambda^{10} + \lambda^{9}
    + \lambda^{8} + \lambda^{5} + \lambda^{4} + \lambda^{1} + 1$\\
14 &$\lambda^{14} + \lambda^{12} + \lambda^{6} + \lambda^{4} + \lambda^{2} +
1$\\
15 &$\lambda^{15} + \lambda^{14} + \lambda^{11} + \lambda^{10}
    + \lambda^{9} + \lambda^{8} + \lambda^{7} + \lambda^{6}
    + \lambda^{3} + \lambda^{2}$\\
16 &$\lambda^{16} + 1$\\
17 &$\lambda^{17} + \lambda^{16} + \lambda^{15} + \lambda^{14}
    + \lambda^{11} + \lambda^{10} + \lambda^{9} + \lambda^{8}
    + \lambda^{7} + \lambda^{6} + \lambda^{3} + \lambda^{2}
    + \lambda^{1} + 1$\\
18 &$\lambda^{18} + \lambda^{16} + \lambda^{14} + \lambda^{12}
    + \lambda^{6} + \lambda^{4}$\\
19 &$\lambda^{19} + \lambda^{18} + \lambda^{13} + \lambda^{12}
    + \lambda^{11} + \lambda^{10} + \lambda^{9} + \lambda^{8}
    + \lambda^{5} + \lambda^{4} + \lambda^{3} + \lambda^{2}
    + \lambda^{1} + 1$\\
20 &$\lambda^{20} + \lambda^{16} + \lambda^{12} + \lambda^{8}
    + \lambda^{4} + 1$\\
21 &$\lambda^{21} + \lambda^{20} + \lambda^{19} + \lambda^{18}
    + \lambda^{17} + \lambda^{16} + \lambda^{11} + \lambda^{10}
    + \lambda^{3} + \lambda^{2}$\\
22 &$\lambda^{22} + \lambda^{20} + \lambda^{10} + \lambda^{8}
    + \lambda^{2} + 1$\\
23 &$\lambda^{23} + \lambda^{22} + \lambda^{19} + \lambda^{18}
    + \lambda^{17} + \lambda^{16} + \lambda^{9} + \lambda^{8}
    + \lambda^{1} + 1$\\
24 &$\lambda^{24} + \lambda^{16}$\\
25 &$\lambda^{25} + \lambda^{24} + \lambda^{23} + \lambda^{22}
    + \lambda^{19} + \lambda^{18} + \lambda^{9} + \lambda^{8}
    + \lambda^{1} + 1$\\
26 &$\lambda^{26} + \lambda^{24} + \lambda^{22} + \lambda^{20}
    + \lambda^{18} + \lambda^{16} + \lambda^{10} + \lambda^{8}
    + \lambda^{2} + 1$\\
27 &$\lambda^{27} + \lambda^{26} + \lambda^{21} + \lambda^{20}
    + \lambda^{17} + \lambda^{16} + \lambda^{11} + \lambda^{10}
    + \lambda^{3} + \lambda^{2}$\\
28 &$\lambda^{28} + \lambda^{24} + \lambda^{12} + \lambda^{8}
    + \lambda^{4} + 1$\\
29 &$\lambda^{29} + \lambda^{28} + \lambda^{27} + \lambda^{26}
    + \lambda^{25} + \lambda^{24} + \lambda^{21} + \lambda^{20}
    + \lambda^{17} + \lambda^{16} + \lambda^{13} + \lambda^{12} $\\
    &$\quad+ \lambda^{11} + \lambda^{10} + \lambda^{9} + \lambda^{8}
    + \lambda^{5} + \lambda^{4} + \lambda^{3} + \lambda^{2}
    + \lambda^{1} + 1$\\
30 &$\lambda^{30} + \lambda^{28} + \lambda^{22} + \lambda^{20}
    + \lambda^{18} + \lambda^{16} + \lambda^{14} + \lambda^{12}
    + \lambda^{6} + \lambda^{4}$\\
31 &$\lambda^{31} + \lambda^{30} + \lambda^{27} + \lambda^{26}
    + \lambda^{25} + \lambda^{24} + \lambda^{23} + \lambda^{22}$\\
   &$\quad + \lambda^{19} + \lambda^{18} + \lambda^{15} + \lambda^{14}
    + \lambda^{11} + \lambda^{10} + \lambda^{9} + \lambda^{8}
    + \lambda^{7} + \lambda^{6}$\\
   &$\quad + \lambda^{3} + \lambda^{2} + \lambda^{1} + 1$\\
32 &$\lambda^{32} + 1$\\
\end{tabular}
\end{table}

\begin{table}
\caption{Period for rule-90 and 150 cylindrical CA.}
\begin{tabular}{rllll}
&\multicolumn{2}{c}{Rule-90}&\multicolumn{2}{c}{Rule-150}\\
$N$&Polynomials&Actual&Polynomials&Actual\\ \tableline
  4&$U^{   4}=0    $&$U^{   2}=0    $&$U^{   4}=I    $&$U^{   2}=I   $\\
  5&$U^{   7}=U    $&$U^{   4}=U    $&$U^{   6}=I    $&$U^{   3}=I   $\\
  6&$U^{   6}=U^{2}$&$U^{   3}=U    $&$U^{   6}=U^{4}$&$U^{   3}=U^{2}$\\
  7&$U^{  15}=U    $&$U^{   8}=U    $&$U^{  14}=I    $&$U^{   7}=I   $\\
  8&$U^{   8}=0    $&$U^{   4}=0    $&$U^{   8}=I    $&$U^{   4}=I   $\\
  9&$U^{  15}=U    $&$U^{   8}=U    $&$U^{  16}=U^{2}$&$U^{   8}=U   $\\
 10&$U^{  14}=U^{2}$&$U^{   7}=U    $&$U^{  12}=I    $&$U^{   6}=I   $\\
 11&$U^{  63}=U    $&$U^{  32}=U    $&$U^{  62}=I    $&$U^{  31}=I   $\\
 12&$U^{  12}=U^{4}$&$U^{   6}=U^{2}$&$U^{  12}=U^{8}$&$U^{   6}=U^{4}$\\
 13&$U^{ 127}=U    $&$U^{  64}=U    $&$U^{  42}=I    $&$U^{  21}=I   $\\
 14&$U^{  30}=U^{2}$&$U^{  15}=U    $&$U^{  28}=I    $&$U^{  14}=I   $\\
 15&$U^{  31}=U    $&$U^{  16}=U    $&$U^{  32}=U^{2}$&$U^{  16}=U   $\\
 16&$U^{  16}=0    $&$U^{   8}=0    $&$U^{  16}=I    $&$U^{   8}=I   $\\
 17&$U^{  31}=U    $&$U^{  16}=U    $&$U^{  30}=I    $&$U^{  15}=I   $\\
 18&$U^{  30}=U^{2}$&$U^{  15}=U    $&$U^{  32}=U^{4}$&$U^{  16}=U^{2}$\\
 19&$U^{1023}=U    $&$U^{ 512}=U    $&$U^{1022}=I    $&$U^{ 511}=I   $\\
 20&$U^{  28}=U^{4}$&$U^{  14}=U^{2}$&$U^{  24}=I    $&$U^{  12}=I   $\\
 21&$U^{ 127}=U    $&$U^{  64}=U    $&$U^{ 128}=U^{2}$&$U^{  64}=U   $\\
 22&$U^{ 126}=U^{2}$&$U^{  63}=U    $&$U^{ 124}=I    $&$U^{  62}=I   $\\
 23&$U^{4095}=U    $&$U^{2048}=U    $&$U^{4094}=I    $&$U^{2047}=I   $\\
 24&$U^{  24}=U^{8}$&$U^{  12}=U^{4}$&$U^{  24}=U^{16}$&$U^{  12}=U^{8}$\\
 25&$U^{2047}=U    $&$U^{1024}=U    $&$U^{2046}=I    $&$U^{1023}=I   $\\
 26&$U^{ 254}=U^{2}$&$U^{ 127}=U    $&$U^{  84}=I    $&$U^{  42}=I   $\\
 27&$U^{1023}=U    $&$U^{ 512}=U    $&$U^{1024}=U^{2}$&$U^{ 512}=U   $\\
 28&$U^{  60}=U^{4}$&$U^{  30}=U^{2}$&$U^{  56}=I    $&$U^{  28}=I   $\\
 29&$U^{32767}=U   $&$U^{16384}=U   $&$U^{32766}=I   $&$U^{16383}=I   $\\
 30&$U^{  62}=U^{2}$&$U^{  31}=U    $&$U^{  64}=U^{4}$&$U^{  32}=U^{2}$\\
 31&$U^{  63}=U    $&$U^{  32}=U    $&$U^{  62}=I    $&$U^{  31}=I   $\\
 32&$U^{  32}=0    $&$U^{  16}=0    $&$U^{  32}=I    $&$U^{  16}=I   $\\
\end{tabular}
\label{actual-period}
\end{table}

\begin{references}
\bibitem[*]{EADDRESS}Electronic address: tadaki@ai.is.saga-u.ac.jp
\bibitem{Neumann}J. von Neumann, in {\it Theory of
Self-Reproducing Automata}, edited by A. W. Burks
(University of Illinois Press, Urbana, 1966).
\bibitem{Wolfram:1}S. Wolfram, Nature {\bf 311}, 419 (1984);
Phys. Rev. Lett. {\bf 55}, 449 (1985).
\bibitem{Wolfram:2}S. Wolfram, Rev. Mod. Phys. {\bf 55}, 601 (1983).
\bibitem{Wolfram:3}S. Wolfram, Physica D{\bf 10}, 1 (1984);
Commun. Math. Phys. {\bf 96}, 15 (1984);
J. Stat. Phys. {\bf 45}, 471 (1986).
\bibitem{Martin}O. Martin, A. M. Odlyzko and S. Wolfram,
Commun. Math. Phys. {\bf 93}, 219 (1984).
\bibitem{CA}H. A. Gutowitz ed., {\it Cellular Automata}
(MIT Press, London, 1991),
J. M. Perdang and A. Lejeune ed., {\it Cellular Automata}
(World Scientific, Singapore, 1993),
N. Boccara, E. Goles, S. Martinez and P. Picco ed.,
{\it Cellular Automata and Cooperative Systems}
(Kluwer Academic, Dorderecht, 1993).
\bibitem{Wolfram:4}S. Wolfram, Adv. Appl. Math. {\bf 7}, 127 (1986).
\bibitem{Jen}E. Jen, Commun. Math. Phys. {\bf 118}, 569 (1988).
\bibitem{Voorhees}B. Voorhees, Physica D{\bf 45}, 26 (1990).
\bibitem{Guan}P. Guan and Y. He, J. Stat. Phys. {\bf 43}, 463 (1986).
\bibitem{Stevens}J. G. Stevens, R. E. Rosensweig and A. E. Cerkanowicz,
J. Stat. Phys. {\bf 73}, 159 (1993).
\bibitem{Tadaki:1}S. Tadaki and S. Matsufuji,
      Prog. Theor. Phys. {\bf 89}, 325 (1993).
\bibitem{Tadaki:2}S. Tadaki, Phys. Rev. E{\bf 49}, 1168 (1994).
\bibitem{num} N. Koblitz, {\it A Course in Number Theory and Cryptography}
      (Springer-Verlag, New York, 1987).
\bibitem{errata}Equation (\ref{REQ90NN}) gives simple relation of
the maximum period and relaxation as $\Pi_{2n+1}=2\Pi_{n}$ and
$\pi_{2n+1}=2\pi_{n}+1$ (there are mistakes in Refs. \cite{Tadaki:1}
and \cite{Tadaki:2}).
\end{references}
\end{document}